\documentclass[pre,showpacs,twocolumn,superscriptaddress,floatfix,amsmath,amssymb]{revtex4} 

\usepackage{graphicx}
\usepackage{times}
\usepackage{graphics}
\usepackage{epsfig}
\usepackage{epstopdf}
\usepackage{graphicx}
\usepackage{amsmath}

\def\el{{\mbox{$\vec{E}$}}}
\def\dip{{\mbox{$\vec{p}$}}}
\def\pd{{\mbox{$\vec{p}_2$}}}

\def\fdep{{\mbox{$\vec{F}_{\rm DEP}$}}}
\def\eunit{{\mbox{$\varepsilon/e\sigma$}}}
\def\tunit{{\mbox{$\varepsilon/k_{\rm B}$}}}

\begin{document}

\title{
Enhanced dielectrophoresis of nanocolloids by dimer formation
}

%

\author{Emppu Salonen}
\affiliation{
  Laboratory of Physics and Helsinki Institute of Physics,
  Helsinki University of Technology,
  Finland
}
\email{emppu.salonen@hut.fi}

\author{Emma Terama}
\affiliation{
  Laboratory of Physics and Helsinki Institute of Physics,
  Helsinki University of Technology,
  Finland
}
\email{emt@fyslab.hut.fi}

\author{Ilpo Vattulainen}
\affiliation{
  Memphys--Center for Biomembrane Physics, Physics Department,
  University of Southern Denmark,
  Odense, Denmark
}

\affiliation{
  Laboratory of Physics and Helsinki Institute of Physics,
  Helsinki University of Technology,
  Finland
}
\affiliation{
  Institute of Physics,
  Tampere University of Technology,
  Finland
}
\email{vattulai@csc.fi}

\author{Mikko Karttunen}
\affiliation{
  Department of Applied Mathematics,
  The University of Western Ontario,
  London, Ontario, Canada
}
\email{mkarttu@uwo.ca}

\begin{abstract}
We investigate the dielectrophoretic motion of charge-neutral, polarizable
nanocolloids through molecular dynamics simulations.  Comparison to analytical
results derived for continuum systems shows that the discrete charge
distributions on the nanocolloids have a significant impact on their coupling
to the external field. Aggregation of nanocolloids leads to enhanced
dielectrophoretic transport, provided that increase in the dipole moment upon
aggregation can overcome the related increase in friction. The dimer
orientation and the exact structure of the nanocolloid charge distribution are
shown to be important in the enhanced transport.
\end{abstract}


\maketitle

\section{Introduction}
Manipulation of microscopic particles with external electric fields,
electrokinetics, has gained considerable attention due to applications
in nanotechnology and biomedical research
\cite{Hughes00,TBJones,MPHughes,Burke04}. One of the central methods
is dielectrophoresis (DEP), the motion of polarizable particles due to
a coupling with a non-uniform electric field \cite{Pohl51}.  Recent
studies have attested the versatility of DEP manipulation with carbon
nanotubes \cite{Kim05}, DNA \cite{Tuukkanen05} and biological
micro-organisms \cite{Markx94}, as well as the assembly of
nanowires \cite{Hermanson01} and biosensor arrays \cite{Gray04}. The
relevance of DEP methods to more complex integrated microscopic
analysis and assembly devices \cite{Reyes02} is evident.

It has been demonstrated that particles down to a few nm in size can be
manipulated with DEP \cite{Bezryadin97,Zheng04}, but it is still not known
what is the minimum size of particles that can be efficiently transported and
trapped \cite{Burke04}. Experiments are hampered by the fact that {\it in
situ} tracking of nanoscale particles is very difficult.  With decreasing
particle size the effect of thermal noise becomes increasingly
important. Furthermore, the motion of polarizable particles in non-uniform
electric fields is inherently a non-equilibrium process, and particles in
close contact may interact via non-additive many-body interactions.  Given
these challenges, it is fair to say that understanding the problem of
nanoscale DEP calls for new theoretical insight.

For an isolated particle much smaller than the characteristic length
of the electric field non-uniformity, the DEP force is given by the
well-known expression \cite{TBJones}
\begin{equation}
\fdep = ({\mathbf \dip} \cdot \nabla) \el,
\label{genforce}
\end{equation}
where \dip{} is the particle dipole moment induced by the field \el.
As the separation between particles decreases, their mutual
interactions lead to modifications in their electric polarizations and
the DEP force affecting them \cite{Huang03,Huang04a,TBJones}. It has
been suggested that such changes result in enhanced electrokinetic
trapping efficiencies at non-dilute conditions
\cite{Muller96}. Controlled particle aggregation may in fact be an
efficient way to overcome the large thermal forces that hinder DEP
trapping and transport. Despite their importance, these issues, to our
knowledge, have not yet been studied neither experimentally nor
theoretically in detail. As for non-equilibrium simulations of DEP,
there is only one recent study focusing on the transport of an
individual nanocolloid \cite{Salonen05}.

In this letter, we consider DEP through theory and modeling and show how
enhanced DEP transport can be obtained by controlled complexation.  We use
molecular dynamics simulations to study the aggregation and subsequent DEP
transport of a dimer comprised of two spherical nanocolloids. We compare the
transport properties of a dimer to those of a single nanocolloid, and analyze
the competing roles of DEP force and friction.

\section{Simulation model}
Our model \cite{Salonen05} is summarized as follows. We start from a charged
spherical macroion with $N_i$ small microions, each of charge $q = +2 e$,
electrostatically bound on the macroion. The macroion charge was set to $Q = -
N_i q$, so that the macroion-microion complex, termed hereafter as {\it
nanocolloid}, is charge-neutral. For excluded volume interactions we used a
shifted repulsive Lennard-Jones potential, $\Phi_\mathrm{WCA} = 4 \varepsilon
[(\sigma/\{ r_{ij} - r_0 \})^{12} - (\sigma / \{ r_{ij} - r_0 \} )^{6} +
1/4]$, where $\varepsilon$ and $\sigma$ are the characteristic energy and
length of the interaction, respectively, and $r_{ij} = \arrowvert \vec{r}_j -
\vec{r}_i \arrowvert$ is the center-to-center distance between particles $i$
and $j$. The potential was truncated at $r_{ij} = r_0 + 2^{1/6} \sigma$. The
hard-core radius $r_0$ describing particle size was assigned a non-zero value
in interactions involving macroions. Electrostatic interactions were
calculated directly from Coulomb's law using a constant relative permittivity
$\epsilon_r$, given by $(4 \pi \epsilon_0 \epsilon_r)^{-1} = 56
\,\varepsilon\sigma/e^2$, where $\epsilon_0$ is the vacuum permittivity.
These types of generic models are physically transparent and have previously
been employed in studies of electrokinetic phenomena in colloidal solutions
\cite{Tanaka02,Linse99,Messina00,Patra03}. Our choice of simulation parameters
can be viewed to model, {\it e.g.}, DEP of reverse micelles \cite{Hakoda96},
reverse phase emulsions \cite{Flores-Rodriguez05}, or high-generation
dendrimers \cite{Jaaskelainen00}.

\begin{figure}[tb]
\includegraphics[width=8cm]{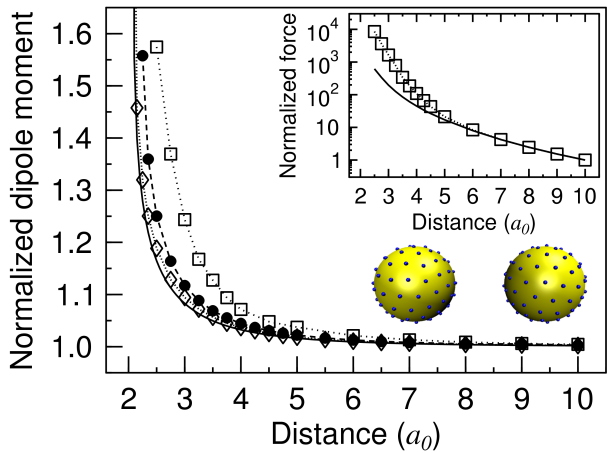}
\caption{
Dipole moments for two nanocolloids aligned parallel to the
electric field, $E_0 = 0.96$\,\eunit, normalized with the dipole 
moment of a single nanocolloid. 
The symbols correspond to $N_i = 10$
(open squares), 100 (filled circles), and 300 (open diamonds). 
The solid line is the analytical continuum 
result \cite{Huang03}. Inset: Total nanocolloid--nanocolloid 
attractive force for $N_i = 10$ (open squares), normalized with 
the value at $D = 10\,a_0$.  The solid line shows the pure dipole-dipole
contribution to the electrostatic force.
}
\label{dipolemoments}
\end{figure}

\section{Results and discussion}
We first studied the influence of nanocolloid interactions on the
dipole moments and the resulting electrostatic forces. The
calculations were carried out in the absence of solvent. 
We considered the two specific cases where the
line joining a pair of symmetric nanocolloids was either perpendicular
or parallel to the direction of a constant electric field $E_0 =
0.96$\,\eunit. We used hard-core radii $r_0$ = 4.5, 16.9,
and 29.6\,$\sigma$ for the macroions, with a fixed microion surface 
density in all cases.  The number of microions then increased with
$r_0$ ($N_i$ = 10, 100, and 300, respectively). The macroions were
set a distance $D$ apart from each other and subsequently held fixed
at these positions. The microions were allowed to move freely, and
their surface configurations were relaxed by slowly quenching the
kinetic energy from the system with the Berendsen thermostat
\cite{Berendsen84}.

Figure~\ref{dipolemoments} shows the nanocolloid dipole moments for a
pair aligned parallel to the electric field as a function of the
nanocolloid separation $D$ (in units of $a_0 = r_0 + \sigma/2$). The
solid line shows the analytical prediction in the continuum limit
\cite{Huang03}. For maximal polarization, we chose the value $b = 1$
in eq.~(7) of ref.~\cite{Huang03}.  For the largest macroion ($N_i =
300$) the enhancement in the dipole moments is in excellent agreement
with the continuum case.  Though, as expected, the agreement becomes
worse with decreasing macroion size. This results from the small
number of microions and pronounced discrete nature of the microion
distribution.  Yet it is remarkable how well the continuum theory
grasps the essential features even just for $N_i \sim 100$.  The
results for the case of nanocolloids aligned perpendicular to the
electric field also followed the same trend (data not shown):
reduction of the dipole moment with decreasing interparticle distance,
as follows from eq.~(8) in ref.~\cite{Huang03}, was the most
significant for $N_i = 10$.  Again, for $N_i = 300$ our simulations
were in an excellent agreement with the analytical result.

The inset in fig.~\ref{dipolemoments} shows the electrostatic force between
the nanocolloids ($N_i = 10$) as a function of distance, together with just
the bare dipole-dipole contribution. As the interparticle distance decreases,
the total electrostatic force reaches a value over an order of magnitude
larger than the dipole-dipole contribution.  By running a series of
consecutive kinetic energy quenching simulations at decreasing values of $D$,
an electrostatically bound dimer was formed.  Some of the microions were
trapped in the region between the macroions. For $E_0 \geq$ 0.96\,\eunit, the
resulting dimer dipole moments were a factor of 3.4\,--\,4.1 higher than for a
single isolated nanocolloid. At $E_0 = 0.68$\,\eunit\, the ratio of dimer and
single nanocolloid dipole moments was only 2.1.  The above highlights the
special nature of nanosized colloids: the smaller the charge, the stronger is
the effect on creating asymmetries as the discrete nature of charge becomes
important \cite{Patra03}.  Here, as the microions are strongly bound, that
leads to non-uniform distribution that enhances the dipole for the pair. 
Note that a naive estimate, based on the notion that polarization is
proportional to volume, would imply only an increase by a factor of two for
the nanocolloid dimer. Here, at the higher values of $E_0$, the increase in
the dipole moments is much larger.

To fully describe the dynamics, we next simulated the DEP motion of a
dimer with $r_0 = 4.5\,\sigma$ and $N_i = 10$.  We chose a spherically
symmetric field, $\el(R) = E_0 (R_0/R)^2 \vec{e}_\mathrm{R}$, where
$R$ is the radial distance from the electric field interaction center,
$R_0 = 1500\,\sigma$ the characteristic length of the field, and
$\vec{e}_\mathrm{R}$ is the radial unit vector. To model DEP transport
in a mesoscopic electric field geometry, with proper changes in the
electric field strength, we employed an additional electric field
system frame of reference and a periodic particle shifting scheme
\cite{Salonen05}. The initial center-of-mass (CM) position of the
dimer was set at $R = R_0$, with the line joining the macroions
parallel to the electric field. A detailed description of our
simulation protocol is given elsewhere \cite{Salonen05}.

The dimer was immersed in an explicit solvent of neutral particles with a
density of $\rho_s = 0.3\,\sigma^{-3}$.  A cubic simulation cell of side
length 45\,$\sigma$ was used. Tests with larger systems showed that this cell
size resulted in minimal finite-size effects. The solvent density in our
coarse-grained model is different from the ones used in simulations of atomic or
molecular liquids, where typically $\rho_s \approx 0.8\,\sigma^{-3}$.  The
solvent density in this letter was also used in the molecular dynamics study
of electrophoresis by Tanaka and Grosberg \cite{Tanaka02}, as well as in our
previous paper on nanocolloid dielectrophoresis \cite{Salonen05}. The
motivation for using $\rho_s = 0.3\,\sigma^{-3}$, is that by reducing the
solvent friction, related to $\rho_s$, it is possible to study the effect of
the spatial variance of the electric field to the DEP transport with
simulations of reasonable length. Second, as we compare the DEP displacements
of single nanocolloids and dimers [see below eq. (\ref{1cprediction})], the
solvent density has no major importance to the ratio of the displacements, and
thus, to the main conclusions of this letter.

For each combination of $E_0$ and temperature $T$ we carried out 50\,--\,160
independent DEP simulations of length 5785\,$\tau$, with the unit of time
given by $\tau = \sigma \sqrt{m/\varepsilon}$.  To give a concrete idea of the
timescales involved, with the central quantities in our model having plausible
nanocolloid values $\sigma$ = 0.2 nm, $m$ = 30 amu (atomic mass unit),
$\varepsilon$ = $k_{\rm B} T_0$, and $T_0$ = 300 K, our simulations would then
correspond to a timescale of $\approx$ 4 ns.  No dimer breakups were observed
in the simulations.

\begin{figure}
\includegraphics[width=\columnwidth]{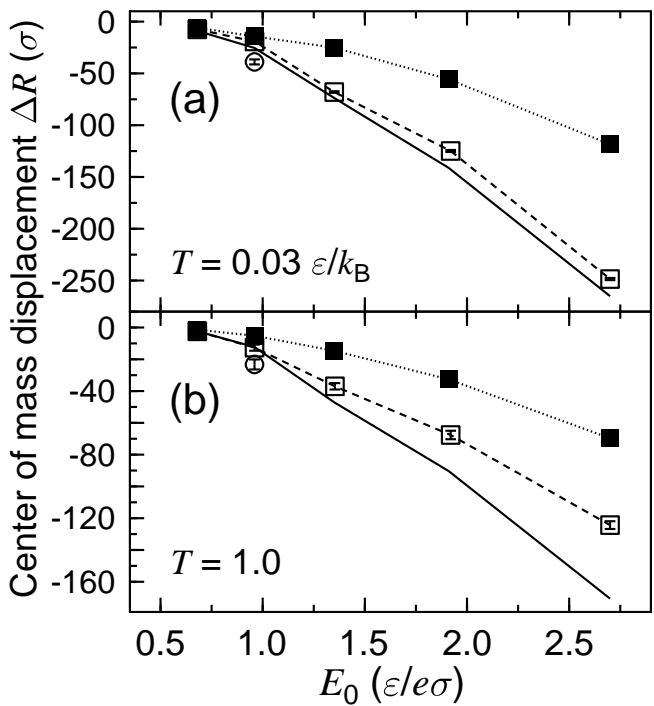}
\caption{
Mean radial DEP displacements at 
(a) $T = 0.03$, and 
(b) 1.0 \tunit{}. 
The open (solid) squares show the data for dimers
(single nanocolloids). The open circles at 
$E_0 = 0.96$\,\eunit{} 
show the data only for the simulations where the dimer assumed
the higher dipole moment charge configuration (see text) for the whole length
of the simulation. The solid lines show the predictions 
 for the dimer case, see eq.~(\ref{1cprediction}).
}
\label{dr}
\end{figure}

Figure~\ref{dr} shows the dimer and single nanocolloid \cite{1cnote}
CM mean radial displacements in the electric field system of reference
for the two extreme temperatures used, $T = 0.03$ and
$1.0\,\varepsilon/k_\mathrm{B}$. At $E_0 \geq$ 1.35\,\eunit{} the
dimer displacements are clearly larger than those of the single
nanocolloid.  For $E_0 = 0.68\,\eunit$ the DEP displacements are the
same within the error bars.

For $E_0 = 0.96$\,\eunit{} it was observed that the dimer polarization
relaxed to two different states, corresponding to two different
configurations of the microions trapped between the macroions.  In
some simulations the dimer assumed a state of very strong polarization
with a dipole moment a factor of 5 higher than that of a single
nanocolloid. The fraction of simulations with this strong dimer
polarization increased with $T$ from 15\% at the lowest $T$ to 33\% at
the highest $T$ used. For the other, more common, polarization state
the dipole moment ratio was only 2.1, {\it i.e.}, a value similar as
for $E_0 = 0.68$\,\eunit.  A more detailed analysis showed that in
some cases the dipole made transitions between the two states.
However, the dimers never retained the original polarized state that
was formed in the absence of the solvent. Similar transitions were not
observed for any other value of $E_0$.

Equation~(\ref{genforce}) shows that the DEP force affecting a dimer is larger
than that on a single nanocolloid. But the frictional force due to the solvent
on the dimer is also larger. Figure~\ref{dr} shows that in our case the DEP
force acting on the dimer outweighs the increase in friction. Whether the DEP
transport of aggregates is enhanced in comparison to single particles {\it in
general} is essentially determined by the relative changes in these two
factors.  An estimate for the ratio of the friction factors of a dimer and a
single spherical nanocolloid, $\xi_2/\xi_1$, can be made from hydrodynamics of
smooth particles. With the long axis of the dimer aligned with the direction
of motion, $\xi_2^{\parallel}/\xi_1 \approx 1.29$ \cite{Swanson78}. For motion
perpendicular to the long axis of the dimer, $\xi_2^{\perp}/\xi_1 \approx
1.43$. This strongly suggests that nanocolloid aggregation enhances DEP
transport, in line with our results in fig.~\ref{dr}; see also discussion
below.

To elaborate this issue, note that it has previously been verified
\cite{Salonen05} that DEP displacements with the range of external
parameters $E_0$ and $T$ used here are linearly proportional to the
mean DEP force.  A simple estimate for the dimer DEP displacement,
$\Delta R_2$, is then obtained from,
\begin{equation}
\Delta R_2 \approx \frac{F_2(R_0)}{F_1(R_0)}
\left( \frac{\xi_1}{\xi_2^{\parallel}} \right) \Delta R_1,
\label{1cprediction}
\end{equation}
where $F_i(R_0)$ is the initial DEP force on the system consisting of
$i$ nanocolloids at $R_0$, and $\Delta R_i$ its displacement at the
corresponding $E_0$.  The resulting values are shown as solid lines in
fig.~\ref{dr}.  The agreement with the simulation data at $T = 0.03$
\tunit{} is good, though eq.~(\ref{1cprediction}) does not properly
account for the spatial dependence of the DEP force.

At increasing temperatures it should be expected that the rotational motion of
the dimer hinders its DEP transport. This is due to two effects. First, for
orientations deviating from full alignment with the direction of motion the
effective friction factor of the dimer increases, see above. Second, the
dipole moment of the rotating dimer changes as the microion distribution
relaxes according to the field.  Although changes in the microion surface
distributions resulted, on the average, in appreciable DEP forces at all
orientations, perpendicular dimer orientation with respect to the direction of
\el{} decreased the coupling strength up to by a factor of two.  Considering
the increased friction (by a factor of $\sim$1.43) together with the
relaxation of the microion distribution, decreasing the DEP force by 50\%, an
increase of \fdep{} by a factor of $\sim$3 is required for enhanced DEP
transport of the nanocolloid dimer when rotations take place.

\begin{figure}
\includegraphics[width=\columnwidth]{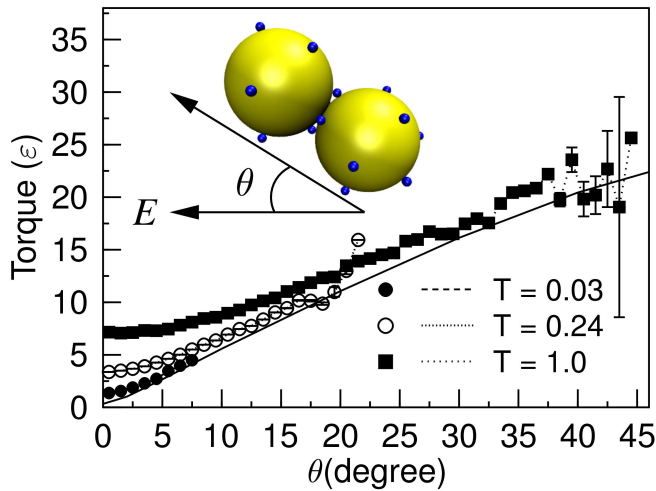}
\caption{
Electrostatic torque $\Gamma$ on the dimer as a function of the
dimer orientation, $E_0$ = 1.91 \eunit. The solid line shows the results
calculated for a relaxed charge distribution.
}
\label{torque}
\end{figure}

Interestingly, the dimer serves as a model to gain insight into the
role of $T$ on rotational DEP motion of rod-like molecules. The
external field counters the dimer rotation by exerting an
electrostatic torque $\vec{\Gamma} = \pd(R,\theta) \times \vec{E}(R),$
where the dipole moment of the dimer, \pd, depends on the radial
position $R$ and the orientation angle $\theta$ with respect to the
electric field. The characteristic time of dimer rotation may be
shorter than the charge relaxation time.  This affects the actual
values the dimer dipole moment assumes at different orientations. To
illustrate this, we show in fig.~\ref{torque} the values of torque
affecting the dimer. Since the electric field is spatially variant,
the data were collected only for $R = 1464$\,--\,1539\,$\sigma$, where
$\vert\el\vert$ did not differ by more than 5\,\% from its value at
$R_0$. We also calculated the torque for fixed dimer orientations in
the absence of solvent.  The resulting charge distributions then
corresponded to the case where the dimer rotation is very slow
compared with the microion relaxation time.

For low $T$ the dimer rotations are slow and result in smaller
absolute values of $\Gamma$. Yet, in comparison to $k_\mathrm{B}T$,
the values of the torque are extremely high and the dimer is
efficiently aligned in the direction of the electric field. As $T$ is
increased, fluctuations in the microion distributions result in higher
values of $\Gamma$ even for small $\theta$. For lower values of $E_0$
full rotations of the dimer were also observed.

\begin{figure}
\includegraphics[width=\columnwidth]{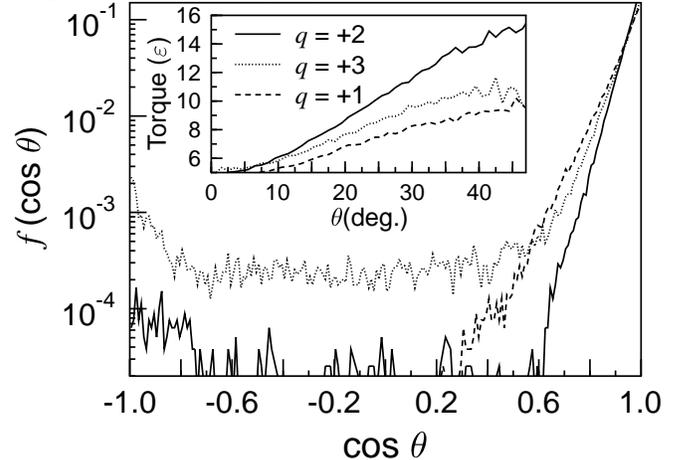}
\caption{
Distributions of the dimer orientations with respect to \el{} in simulations
with $E_0$ = 1.35\,\eunit{} and $T$ = 1.0\,\tunit.  Inset: values of the
electrostatic torque $\Gamma$ for different microion valences.
}
\label{orientation}
\end{figure}

In addition to the case $q$ = $+2e$ and $N_i$ = 10, we further
explored the effects of the microion valence by using two other
microion configurations, namely $q$ = $+1e$, $N_i$ = 20 and $q$ =
$+3e$, $N_i$ = 7. The orientational distributions of the dimers are
shown in fig. \ref{orientation}. Note that the distributions are
biased toward $\cos \theta$ = 1 due to the initial orientation of the
dimer in the simulations. At $E_0$ = 1.35\,\eunit{} and $T$ =
1.0\,\tunit{} the DEP displacements for $q$ = $+1e$ and $+3e$ were
19\% and 16\% smaller, respectively, than for $q$ = $+2e$. These
differences were statistically significant with respect to the margins
of error. The increased dimer rotation for the two extreme values of
valence is also indicated by the smaller values of the electrostatic
torque affecting the dimer (cf. inset of fig. \ref{orientation}).
These results demonstrate that the microion valence could in fact be a
control parameter for nanocolloid DEP motion.

So far rotational effects under DEP have been considered only for much
larger objects with large aspect ratios, such as carbon nanotubes
\cite{Kim05} and DNA \cite{Germishuizen05}.  In these cases large
torques are induced, aligning the macromolecules efficiently along the
electric field. However, in our case even small perturbations of the
microscopic charge distribution due to thermal fluctuations can result
in considerable changes in $\Gamma(R,\theta)$, see fig.~\ref{torque}.
The rotations could lead to non-negligible hydrodynamic coupling
between DEP-manipulated dimers or larger aggregates. Though, as
previously mentioned, our modeling can be viewed to correspond to a
timescale of a few ns. Since we model a dilute system, possible
hydrodynamic effects between rotating dimers are thus not relevant
here.

\section{Conclusions}
We have shown how the proximity of nanocolloids with discrete charge
distributions increases their DEP coupling. By forming nanocolloid
dimers, enhanced transport can be obtained provided that the increase
in the dipole moment overcomes the increase in the friction.  We have
found the rotational motion of dimers to hinder the transport due to
smaller DEP coupling and larger friction. The microion valence was
shown to affect dimer orientation and transport. Our results indicate
that controlled aggregation can be an efficient way of overcoming
thermal forces and obtaining high-precision DEP particle manipulation
at the nanoscale.

\acknowledgments
We thank K. W. Yu and J. P. Huang for valuable
discussions.  Support from the Academy of Finland, Emil Aaltonen foundation
and NSERC of Canada is acknowledged.  We thank the Finnish IT Center (CSC),
and the DCSC Center at the University of Southern Denmark for computing time.

\end{document}